# Oxygen-vacancy-related relaxation and scaling behaviors of $Bi_{0.9}La_{0.1}Fe_{0.98}Mg_{0.02}O_3$ ferroelectric thin film

Qingqing Ke[1], Xiaojie Lou[1], Yang Wang[2] and John Wang[1, a)]

[1] Department of Materials Science and Engineering, National University of Singapore, Singapore 117574

[2] School of Materials Science and Engineering, Nanyang Technological University, 50 Nanyang Singapore 639798

a) Author to whom corresponding should be addressed. Electronic mail: msewangj@nus.edu.sg

**ABSTRACT**

Oxygen-vacancies-related dielectric relaxation and scaling behaviors of $Bi_{0.9}La_{0.1}Fe_{0.98}Mg_{0.02}O_3$ (BLFM) thin film have been investigated by temperature-dependent impedance spectroscopy from 40 $^oC$ up to 200 $^oC$. We found that hopping electrons and single-charged oxygen vacancies ($V_O^{\bullet}$) coexist in the BLFM thin film and make contribution to dielectric response of grain and grain boundary respectively. The activation energy for $V_O^{\bullet}$ is shown to be 0.94 eV in the whole temperature range investigated, whereas the distinct activation energies for electrons are 0.136 eV below 110$^oC$ and 0.239 eV above 110$^oC$ in association with hopping along the $Fe^{2+}$- $V_O^{\bullet}$-$Fe^{3+}$ chain and hopping between $Fe^{2+}$-$Fe^{3+}$, respectively, indicating different hopping processes for electrons. Moreover, it has been found that hopping electrons is in form of long rang movement, while localized and long range movement of oxygen vacancies coexist in BLFM film. The Cole-Cole plots in modulus formalism show a poly-dispersive nature of relaxation for oxygen vacancies and a unique relaxation time for hopping electrons. The scaling behavior of modulus spectra further suggests that the distribution of relaxation times for oxygen vacancies is temperature independent.

## I. INTRODUCTION

The coexistence of (anti)ferroelectric and (anti)ferromagnetic orders in multiferroic thin films has attracted a considerable amount of interests due to the potential applications in the wide spectrum of spintronics, data storage, and microelectromechanical devices.[1-3] Among the very limited number of known multiferroic materials at room temperature, $BiFeO_3$ (BFO) has drawn intensive attention due to its high ferroelectric transition ($T_C$~1103 K) and antiferromagnetic transition ($T_N$~643K) temperatures.[4, 5] In particular, a giant polarization $P_r$ at the level of 100 $\mu C/cm^2$ has been observed in BFO thin films and single crystals,[6, 7] demonstrating that they are ideal candidate materials for lead-free ferroelectric applications. Unfortunately, the leakage current in BFO-based thin films is normally very high, which has been one of the main drawbacks impeding their widely expected applications. Generally, the high leakage current has been ascribed to the existence of oxygen vacancies and $Fe^{2+}$, both of which can form impurity energy levels in the band gap of BFO.[8] Recent researches have shown that oxygen vacancies, rather than $Fe^{2+}$, make more contribution towards the high leakage current,[9-12] although it has been argued that oxygen vacancies can decrease the leakage current by forming complex defect orders such as ${Zn^{2+}}_{Fe} - V_O^{\bullet\bullet}$,[13] which could dramatically enhance the endurance voltage.

Oxygen vacancies in BFO-based thin films are also believed to strongly affect their fatigue behavior. According to the model proposed by Dawber and Scott,[14, 15] it is likely that oxygen vacancies migrate towards the electrode interface under ac electric field, where they aggregate and form two-dimensional arrays. The structures formed by oxygen vacancies are supposed to pin the domain walls, thus dramatically decreasing the remnant polarization. Interestingly, as has been reported by Yang *et al.*,[16] the migration of oxygen vacancies in a Ca-doped BFO thin film can form a p-n junction under an appropriate external electrical field. Therefore, the application of an electric field can be employed to modulate the insulator-conductor transition by creating, erasing and inverting the p-n junction.

Since oxygen vacancies play a predominant role in determining the electrical behavior of BFO-based thin films, it would be of considerable interest to conduct an investigation on their relaxation behavior, which would provide insights into the migration kinetics of charge defects, and therefore into a better understanding of the interplay between defects and the external field. Although some dielectric studies have been carried out on BFO thin films[12, 17, 18] in the past several years, few has been done with the understanding of exactly how oxygen vacancies and their interactions have contributed towards the dielectric behavior of BFO-based thin films, where hopping electrons are usually inevitably involved.

In the present work, we have investigated the oxygen-vacancy- related relaxation and scaling behaviors of La, Mg co-doped BFO (BLFM) thin film deposited by RF-magnetron sputtering through conducting systematic complex impedance spectra (CIS) at varying temperatures. We found that single charged oxygen vacancies ($V_O^{\bullet}$) and hopping electrons coexist in the BLFM thin film, where they occur in two different relaxation regimes. The physical nature of relaxation process corresponding to oxygen vacancies and hopping electrons is discussed.

## II. EXPERIMENTAL PROCEDURE

The BLFM thin film was prepared by RF-magnetron sputtering. Before deposition of the BLFM film, a thin $SrRuO_3$ (SRO) buffer layer (~80 nm in thickness) was deposited on Pt/$TiO_2$ /$SiO_2$ /Si at 600°C by RF-sputtering. The BLFM film of 300nm in thickness was then deposited on the top of the SRO layer at 620°C, using the BLFM ceramic target with 10% excess $Bi_2O_3$. The ceramic target was prepared by calcination of an oxide mixture of $Bi_2O_3$, $La_2O_3$, $Fe_2O_3$ and MgO at 750°C, followed by sintering at 820°C in air. The addition of 10% excess $Bi_2O_3$ was used to compensate the loss of Bi during the subsequent thermal treatment process. Phase analyses of the BLFM film were carried out by X-ray diffraction (Bruker D8 Advanced XRD, Bruker AXS Inc., Madison, WI, Cu $K\alpha$ ), which confirmed that the film has a pure perovskite phase without secondary phases. Prior to electrical measurements, Au dots of 200μm in diameter were sputtered on the film using a shallow mask to form top electrodes. The temperature dependent complex impedance spectroscopy (CIS) studies were conducted using a Solartron impedance analyzer. The impedance measurements were carried out from 40 °C to 200 °C in the frequency range of 0.1 Hz to $10^6$ Hz.

## III. RESULTS AND DISCUSSION

### A. Defect structure

In stoichiometric BLFM thin films, oxygen vacancies can be introduced via doping of Mg. However, oxygen vacancies and other charge carriers (e.g., electrons and holes) can also be generated steadily at various stages of calcination and sintering at high temperatures during the target preparation, and in particular during the film deposition process at low partial pressure via rf sputtering. All the defects involved in the film will influence the dielectric relaxation behavior, therefore, various defects as well as their formation processes should be considered. These processes can be described by the $Kr\ddot{o}ger-Vink$ notations as follows.

In a BLFM film, oxygen vacancies are introduced to compensate for the charge balance upon the doping of acceptor Mg, according to

$$2MgO \rightarrow 2Mg'_{Fe} + 2O^{\times} + V_O^{\bullet\bullet} \qquad (1)$$

where $V_O^{\bullet\bullet}$ represents an oxygen vacancy with two positive charges. That is, when two $Mg^{2+}$ replace two $Fe^{3+}$, a $V_O^{\bullet\bullet}$ is introduced to maintain the charge neutrality. Moreover, it is well documented that doping a small amount of Lanthanum helps to stabilize the perovskite structure of BFO phase and suppress the volatilization of Bi and O.[19, 20] Therefore, the process described in Eq. (1) is the predominant way of creating oxygen vacancies.

During the sputtering process at high temperatures (e.g., 620°C) and low oxygen partial pressure (~$10^{-2}$ mbar), oxygen loss is further accelerated, leading to the formation of more oxygen vacancies, which can be expressed as

$$O_O^{\times} \to V_O + \frac{1}{2} O_2 \qquad (2)$$

$$V_O \to V_O^{\bullet} + e' \qquad (3)$$

$$V_O^{\bullet} \to V_O^{\bullet\bullet} + e' \qquad (4)$$

Conduction electrons can be released from neutralized oxygen vacancies by the first-ionized step and the second-ionized step respectively, as shown in Eq. (3) and (4). Then they may transport in an oxidation-reduction process between $Fe^{2+}$ and $Fe^{3+}$ in the form of

$$Fe'_{Fe} \to Fe_{Fe}^{\times} + e' \qquad (5)$$

Besides their association with Fe, these electrons can be weekly bonded to oxygen vacancies which can form a shallow level to trap electrons.[21] Therefore, the localized electrons can transport along the chain between oxygen vacancies and neighboring transition metals, especially under a highly reducing condition, in form of $Fe^{2+}$-$V_O^{\bullet}$-$Fe^{3+}$.[22] These processes can co-exist and make it difficulty to detect the exact locations of electrons, which are dependent on the local structure and temperature.[23]

During the cooling process from high deposition temperature (i.e., 620$^{o}$C in the present work) to room temperature in air, the following re-oxidation can take place:

$$2V_O^{\bullet} + O_2 + 2e' \to 2O_O^{\times} \qquad (6)$$

This process will primarily occur at the grain boundaries, therefore, leading to the formation of highly conductive grains with high oxygen-deficiency and insulating grain boundary.[24]

## B. Relaxation behavior

The frequency dependence of the real ($\varepsilon'$) and imaginary ($\varepsilon''$) parts of dielectric permittivity for the BLFM film at various temperatures are shown in Fig. 1(a) and (b), respectively. For $\varepsilon'$, a plateau of around 400 appears below $10^4$ Hz. Such step-like behavior is quite similar to those in the giant dielectric constant materials such as $CaCu_3Ti_4O_{12}$(CCTO).[25] The magnitude of $\varepsilon'$ decreases with increasing frequencies, demonstrating a typical characteristic of ferroelectric thin films.[26] In Fig. 1(b), the loss peak is centered in the dispersion region of $\varepsilon'$ and shifts towards higher frequencies with increasing temperature, showing a thermally activated process in the BLFM film.

In order to analyze the type of defects involved, one can calculate the activation energies by fitting the loss-peak frequencies using the Arrhenius law:

$$\omega_m = \omega_o \exp(-\frac{E_a}{K_B T}) \qquad (7)$$

where $\omega_m$ is the frequency corresponding to the loss peak, $\omega_o$ is the pre-exponential factor, $E_a$ is the activation energy for relaxation, $K_B$ is the Boltzmann constant and $T$ is absolute temperature. The plot of $\ln\omega_m$ (vs) $1000/T$ is presented in inset of Fig. 1(b), which can be clearly divided into two sections with fitting activation energies of 0.136 eV at low temperatures (<110$^o$C) and 0.239 eV at high temperatures (>110$^o$C), respectively.

While the fitting activation energy (0.239 eV) for the high temperature region is consistent with the previously reported value of 0.325 eV for the relaxation in $BiFeO_3$ ceramics, explained as a two-site electron hopping between $Fe^{2+}$ and $Fe^{3+}$,[27] the low temperature one (0.136 eV) is much smaller, suggesting a different process of electron movement from the directly hopping between $Fe^{2+}$ to $Fe^{3+}$. Considering the rather low first ionization energy of 0.1 eV for oxygen vacancy,[28] we believe that the electron hopping through the $Fe^{2+}$-$V_O^{\bullet}$-$Fe^{3+}$ with a relatively smaller activation energy dominates in the low temperature region. The bridging effect of first-ionized oxygen vacancies ($V_O^{\bullet}$) can dramatically lower the energy barrier between $Fe^{2+}$ and $Fe^{3+}$ for electron hopping. Therefore, we conclude that electron hopping in our film takes two pathways: direct hopping between $Fe^{2+}$ to $Fe^{3+}$ when T > 110°C, and jumping through the bridging oxygen vacancy between $Fe^{2+}$ and $Fe^{3+}$ in the temperature range of 40°C to 110°C.

A detailed analysis on the correlation between $\varepsilon'$ and $\varepsilon''$, as illustrated in inset of Fig. 1(a), confirms that the peak frequency of $\varepsilon''$ coincides with the frequency corresponding to the maximum slope of $\varepsilon'$. This correlation is described by the following relationship, in limit of $\tau_{\min} < 1/\omega < \tau_{\max}$.[29]

$$\varepsilon''(\omega,T) = \frac{\pi}{2}\frac{\partial\varepsilon'(\omega,T)}{\partial(\ln\omega)} \tag{8}$$

The comparison of the measured $\varepsilon''$ of the BLFM film with that calculated from Eq. (8) is shown in Fig. 2. A reasonable match of peak positions can be observed between the experimental and fitting curve. One also notices that the fitting one is of a higher magnitude and has a sharper peak in comparison with the experimental data. These discrepancies may result from the mono-dispersive nature of the relaxation process for hopping electrons in our film, which is different from the assumption for deriving Eq. (8): that is, a range of relaxation times $\tau_{\min} < 1/\omega < \tau_{\max}$. To confirm this mono-dispersive nature, complex Cole-Cole plot between $\varepsilon'$ and $\varepsilon''$ is employed, which can be described by the following empirical relation:[30]

$$\varepsilon^* = \varepsilon' - j\varepsilon'' = \varepsilon_\infty + (\varepsilon_s - \varepsilon_\infty)/[1 + (j\omega\tau)^{1-\alpha}] \quad (9)$$

where $\varepsilon_s$ and $\varepsilon_\infty$ are the static and high frequency dielectric permittivity, respectively, $\tau$ is the relaxation time, and $\alpha$ denotes the distribution of relaxation time and the value of $\alpha\pi/2$ is the angle between the real axis and the line to the circle centre from the high-frequency intercept.[31] For a mono-dispersive relaxation process, one expects the circle center is located exactly on the $\varepsilon'$-axis, whereas for a poly-dispersive process, the circle center will be located below the $\varepsilon'$-axis.[26] Fig. 3 shows two representative Cole-Cole plots by fitting the data obtained at 40°C and 120°C, respectively, using Eq. (9). We can see that the circle centers are located exactly on the $\varepsilon'$-axis confirming the mono-dispersive nature of the relaxation processes for hopping electrons.

Since the relaxation peaks at low frequency ($<1\times10^4$ Hz) can be obscured by conductivity loss, investigation into the electric modulus ($M^*$) was conducted. By definition, $M^*$ is defined as

$$M^* = M' + jM'' = \frac{1}{\varepsilon^*} = \frac{\varepsilon'}{|\varepsilon|^2} + j\frac{\varepsilon''}{|\varepsilon|^2} \quad (10)$$

Note that modulus is reciprocal of complex permittivity. Therefore, the more the conductivity loss contributes to the dielectric permittivity, the less conductivity loss affects the modulus.

$M'$ and $M''$ as a function of frequency in the temperatures range of 40 °C to 200 °C are plotted in Fig. 4. One can see that two plateaus appear in the $M'$ curve and the value of $M'$ increase with increasing frequency. Two $M''$ peaks could also be clearly observed with the one at the low frequency region (below $10^4$ Hz) labeled as Peak I and the other at high frequency region (above $10^4$ Hz) labeled as Peak II, respectively. The positions of both peaks shift towards higher frequencies with increasing temperature. Peak I disperses much stronger than peak II, indicating two different thermal activation processes exist in the BLFM film.

Since $M^*$ shares the same mechanism as $\varepsilon^*$, the latter has been used to analyzed the relaxation behavior at high frequency (>$10^4$ Hz). Here we only use the modulus formalism to study the relaxation mechanism for peak I, by using the Arrhenius law. According to the fitting results of peak I as shown in inset of Fig. 4, the activation energy for the relaxation process is 0.94 eV, comparable to the activation energy of 0.84 eV measured for the diffusion of oxygen ions in titanate-based bulk materials such as $BaTiO_3$ and $Pb(Zr/Ti)O_3$.[32, 33] Hence, the relaxation process at low frequencies can be ascribed to the diffusion of oxygen vacancies, either $V_O^{\bullet}$ or $V_O^{\bullet\bullet}$ . We ascribe the defects responsible for the low frequency relaxation to $V_O^{\bullet}$, although our film has a lower activation energy of 0.94 eV than that (~1.54 eV) previously reported for $V_O^{\bullet}$ by Deng *et al.*[34] The possible reason is that doping Mg into $Bi_{0.9}La_{0.1}FeO_3$ is supposed to dramatically increase the concentration of oxygen vacancies, therefore lowering the activation energy.[35]

The complex impedance data obtained for various temperatures ranging from 40$^{o}$C to 180$^{o}$C are shown as Cole-Cole diagrams in Fig. 5. At low temperatures, the straight lines with big slope indicate that our film is highly insulating. However, the impedance decreases with increasing temperature as expected. From inset of Fig. 5, two poorly resolved semicircular arcs can be observed, which are attributed to the responses of grains and that of grain boundaries[17], respectively. Obviously, the small arc at high frequencies tends to be obscured by the large arc at low frequencies, due to the large difference in the magnitudes of resistance between grain and grain boundary.

In order to obtain reliable values for the resistance of grain and grain boundary and establish a connection between microstructure and electrical properties, the equivalent circuit model shown in Fig. 6 is employed, which is based on the brick layer model.[36] In Fig. 6, $R_s$ is the resistance of lead used in the equipment, $C_g$ denotes the capacitance related to domain and dipole reorientation in gain, $R_g$ is the resistance associated with grain, $C_{gb}$ is the capacitance related to grain boundary layer, $R_{gb}$ denotes the resistance across the grain boundary layer, and CPE is a constant phase element indicating the departure from ideal Debye-like model. The CPE admittance is $Y_{CPE} = A_o(j\omega)^n = A\omega^n + jB\omega^n$, with,

$$A = A_o \cos(\frac{n\pi}{2}), \quad B = A_o \sin(\frac{n\pi}{2}) \qquad (11)$$

where $A_0$ and n are parameters depending on temperature only, $A_0$ confines the magnitude of the dispersion and $0 \le n \le 1$. The parameter of $n$ is equal to 1 for ideal capacitor and equal to 0 for ideal resistor.[36]

Our data were then fitted using software ZSIMP WIN version 2 by assuming the equivalent circuit discussed above. $R_g$ and $R_{gb}$ thus obtained are summarized in Table I. The fitting results indeed show that $R_{gb}$ is much higher than $R_g$, due to the lower concentration of oxygen vacancies and trapped electrons in grain boundaries after re-oxidation process shown in Eq. (6). Fig. 7 shows the dc conductivity $\sigma_g$ and $\sigma_{gb}$ ( $\sigma_g \propto 1/R_g$ and $\sigma_{gb} \propto 1/R_{gb}$ ), plotted against the reciprocal temperature in the Arrhenius format. One can see that both $\sigma_g$ and $\sigma_{gb}$ obey the Arrhenius law with respective activation energy of 0.15 eV and 0.91 eV. The dc conductivity activation energies are close to the values obtained from the formalisms of $M''$ and $\varepsilon''$. It further shows that the relaxation process in grain arises largely from hopping electrons and that in grain boundary from the movement of oxygen vacancies.

In the previous discussion, the dielectric relaxation has been analyzed in the formalisms of $M^*$ and $\varepsilon^*$, where localized movement of carriers is dominant. In the case of long range movement, the resistive and conductive behaviors are often analyzed by $Z^*$ and $Y^*(1/Z^*)$.[36] Therefore, the combined plot of $M''$(or $\varepsilon''$) and $Z''$ (or $Y''$) versus frequency is able to distinguish whether the short range or long range movement of charge carries is dominant in a relaxation process. The separation of peak frequencies between $M''$(or $\varepsilon''$) and $Z''$ (or $Y''$) indicates that the relaxation process is dominated by the short range movement of charge carriers and departs from an ideal Debye-like behavior, while the frequencies coincidence suggests that long rang movement of charge carriers is dominant.[36, 37] The normalized functions of $Z''/Z''_{\max}$ and $M''/M''_{\max}$ for peak I and those of $Y''/Y''_{\max}$ and $M''/M''_{\max}$ for peak II at a typical temperature of 80 $^{o}$C are shown in Fig. 8(a) and (b), respectively. The slight mismatch in peak frequency for peak I associated with the movement of oxygen vacancies indicates the occurrence of localized movement of oxygen vacancies and justifies the presence of CPE phase in the equivalent circuit model.[38, 39] For peak II, the overlapping of the peak frequencies between the normalized $M''$ and $Y''$ suggests that electron hopping is a long-range movement, confirming our previous argument that the relaxation of hopping electrons follows an ideal Debye-like behavior.

### C. Scaling behavior

The Cole-Cole plot in formalism of impedance cannot clearly discern responses of grain and grain boundary, as shown in Fig. 5. Upon being plotted into modulus formalism as shown in Fig. 9 at the representative temperatures of 40$^{o}$C and 120$^{o}$C, two distinct semi-circles are identified, which are related to grain and grain boundary responses as marked in the plot. For the grain response, the circular centers are located exactly on $M'$-axis (i.e., $\alpha = 0$), even when temperature is increased by 80$^{o}$C, indicating an ideal Debye relaxation with an unique relaxation time. This is consistent with previous analysis using the formalism of $\varepsilon^*$. However, for the grain boundary response, the circle centers are located below the $M'$-axis and $\alpha$ in the range of [0.086, 0.109] decreases with increasing temperature, demonstrating that relaxation time $\tau$ is not single-valued but distributed continuously or discretely around the mean time $\tau_m = 1/\omega_m$.

The Cole-Cole plot in the modulus formalism justifies a poly-dispersive nature for the dielectric relaxation at low frequencies. However, the small variation of $\alpha$ cannot be used to confirm whether the relaxation time is temperature dependant or not, because of the uncertainties in fitting circles.[26] Therefore, we plotted the $M''(\omega, T)$ in scaled coordinates, i.e., $M''_{(\omega,T)}/M''_{\max}$ versus log ($\omega/\omega_m$), where $\omega_m$ the loss peak frequency. If all the modulus loss profiles are collapsed into one master curve, it suggests that the relaxation time is temperature independent.[26] As shown in Fig. 10, all peaks related to the oxygen vacancies responses indeed collapse into one master curve and almost perfectly overlap at different temperatures ranging from 40°C up to 200°C. It suggests that the dynamic processes of oxygen vacancies occurring at different time scales exhibit the same activation energy and that the distribution of the relaxation times is temperature independent.[40]

## IV. CONCLUSIONS

A thorough investigation has been made into the relaxation and scaling behaviors of the polycrystalline $Bi_{0.9}La_{0.1}Fe_{0.98}Mg_{0.02}O_3$ (BLFM) ferroelectric thin film in the temperature range of 40°C to 200°C. The temperature-dependant modulus and dielectric spectra show that single-charged oxygen vacances ($V_O^{\bullet}$) and hopping electrons coexist in the BLFM thin film. An activation energy of 0.94 eV is extracted and has been attributed to $V_O^{\bullet}$ in the temperatures range investigated. However, two different electron hopping processes with the activation energy of 0.136 eV below 110°C and 0.239 eV above 110°C are observed, and are ascribed to hopping between $Fe^{2+}$-$V_O^{\bullet}$-$Fe^{3+}$ and along $Fe^{2+}$-$Fe^{3+}$, respectively. The Cole-Cole plots in the modulus formalism show a poly-dispersive nature for the relaxation related to $V_O^{\bullet}$ and a unique relaxation time for hopping electrons. The scaling behavior of modulus spectra demonstrates that the distribution of the relaxation times for $V_O^{\bullet}$ is temperature independent.

**TABLE I.** Summary of the electrical parameters obtained from measured data at various temperatures for BLFM film using the equivalent circuit model

| Temperatre [$^{o}$C] | Rs [$\Omega$] | $R_g$ [$\Omega$] | $C_g$ [F] | $R_{gb}$ [$\Omega$] | $C_{gb}$ [F] | $CPE_{gb}$ [F] | n |
|---|---|---|---|---|---|---|---|
| 40 | 327.8 | 31373 | $1.87\times10^{-10}$ | $3.86\times10^{9}$ | $1.95\times10^{-10}$ | $3.21\times10^{-11}$ | 0.66152 |
| 60 | 321.9 | 24996 | $1.89\times10^{-10}$ | $6.95\times10^{8}$ | $2.02\times10^{-10}$ | $1.60\times10^{-10}$ | 0.4761 |
| 100 | 305.5 | 13921 | $2.04\times10^{-10}$ | $2.49\times10^{7}$ | $2.12\times10^{-10}$ | $7.28\times10^{-10}$ | 0.4949 |
| 120 | 299.1 | 9660 | $2.09\times10^{-10}$ | $5.76\times10^{6}$ | $2.20\times10^{-10}$ | $2.61\times10^{-9}$ | 0.4949 |
| 140 | 279.6 | 6196 | $2.19\times10^{-10}$ | $1.46\times10^{6}$ | $2.26\times10^{-10}$ | $5.10\times10^{-9}$ | 0.42562 |
| 160 | 227.5 | 5204 | $2.27\times10^{-10}$ | $4.01\times10^{5}$ | $2.25\times10^{-10}$ | $3.82\times10^{-9}$ | 0.43578 |
| 180 | 205.1 | 4489 | $2.37\times10^{-10}$ | $1.27\times10^{5}$ | $2.38\times10^{-10}$ | $1.48\times10^{-8}$ | 0.52737 |

# Captions for Figures:

FIG. 1. Frequency dependence of $\varepsilon'$ (a) and $\varepsilon''$ (b) of BLFM film at various temperatures. Inset of Fig. 1(a) shows the frequency dependence of $\varepsilon'$ and $\varepsilon''$ at 40$^{o}$C. Inset of Fig. 1(b) shows the temperature dependence of hopping frequency obtained from the $\varepsilon''$ spectra.

FIG. 2. Comparison of the measured $\varepsilon''$ of BLFM with that calculated from Eq. (8) for data obtained at the temperature of 40$^{o}$C.

FIG. 3. Cole-Cole plots at temperatures of 40$^{o}$C and 120$^{o}$C for BLFM film. The arrow shows the direction of increasing frequency.

FIG. 4. Frequency dependence of $M'$ (a) and $M''$ (b) of the BLFM film at various temperatures. Inset of Fig. 4(b) shows the temperature dependence of hopping frequency obtained from $M''$ spectra for peak I.

FIG. 5. Complex impedance plots, $Z''$ vs $Z'$ for the BLFM thin film, at different temperatures ranging from 40$^{o}$C to 180$^{o}$C. The inset one shows the enlarged impedance plot of data obtained at 40$^{o}$C. The arrow shows the direction of increasing frequency.

FIG. 6. The equivalents circuit model used here in associated with brick layer model.

FIG. 7. Temperature dependence of the dc conductivities for grain and grain boundary

obtained from the fitting results of equivalent circuit.

FIG. 8. Normalized imaginary parts of electric modulus $M''/M''_{max}$, and impedance $Z''/Z''_{max}$ corresponding to oxygen vacancies as functions of frequency at 80°C (a). Normalized imaginary parts of modulus $M''/M''_{max}$ and admittance $Y''/Y''_{max}$ corresponding to hopping electrons as functions of frequency at 80°C (b).

FIG. 9. Complex modulus plots of the BLFM film at 40°C and 120°C. The arrow shows the direction of increasing frequency.

FIG. 10. Scaling behavior of $M''$ at various temperatures for oxygen vacancies in BLFM thin film.

FIG. 1.

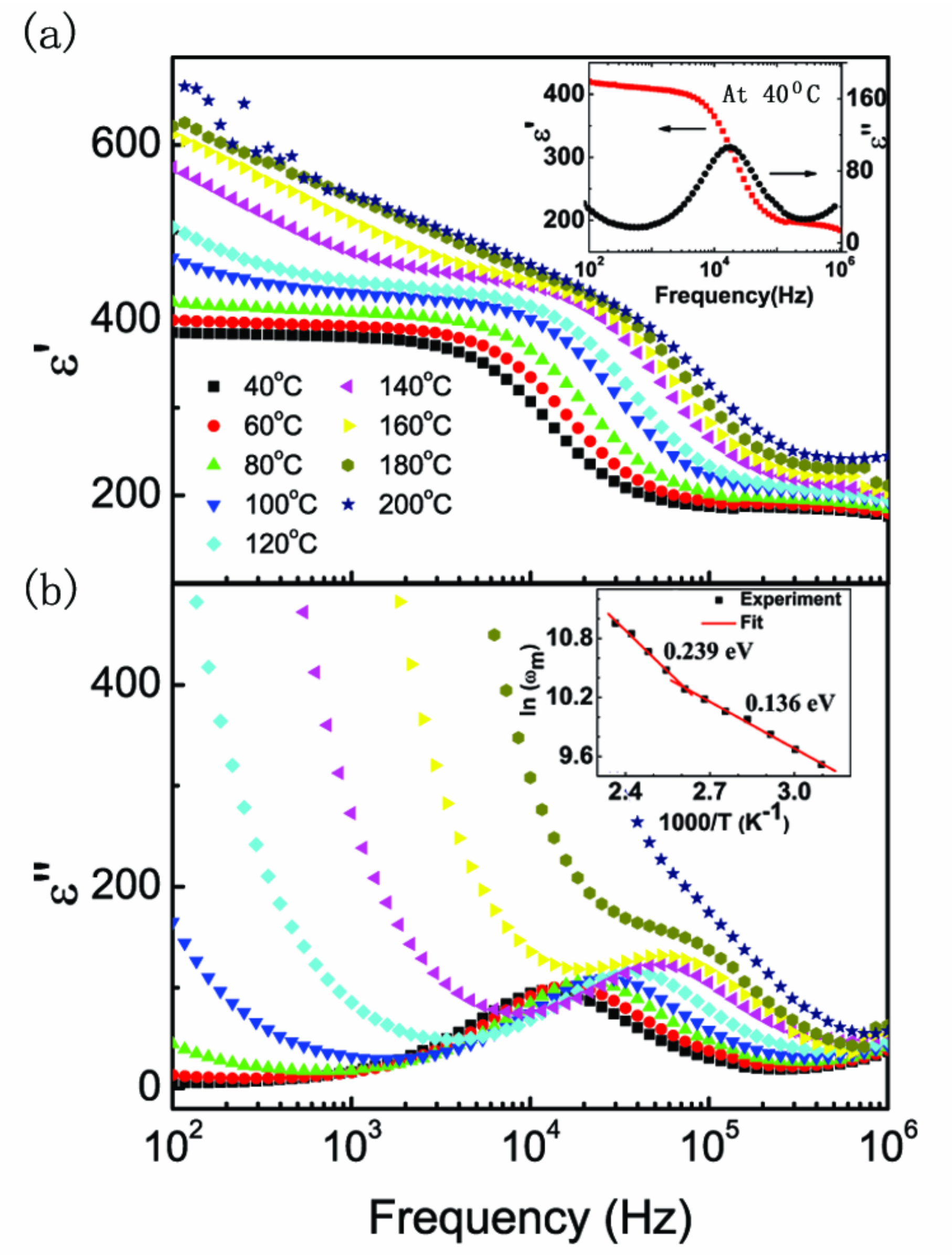

**FIG. 2.**

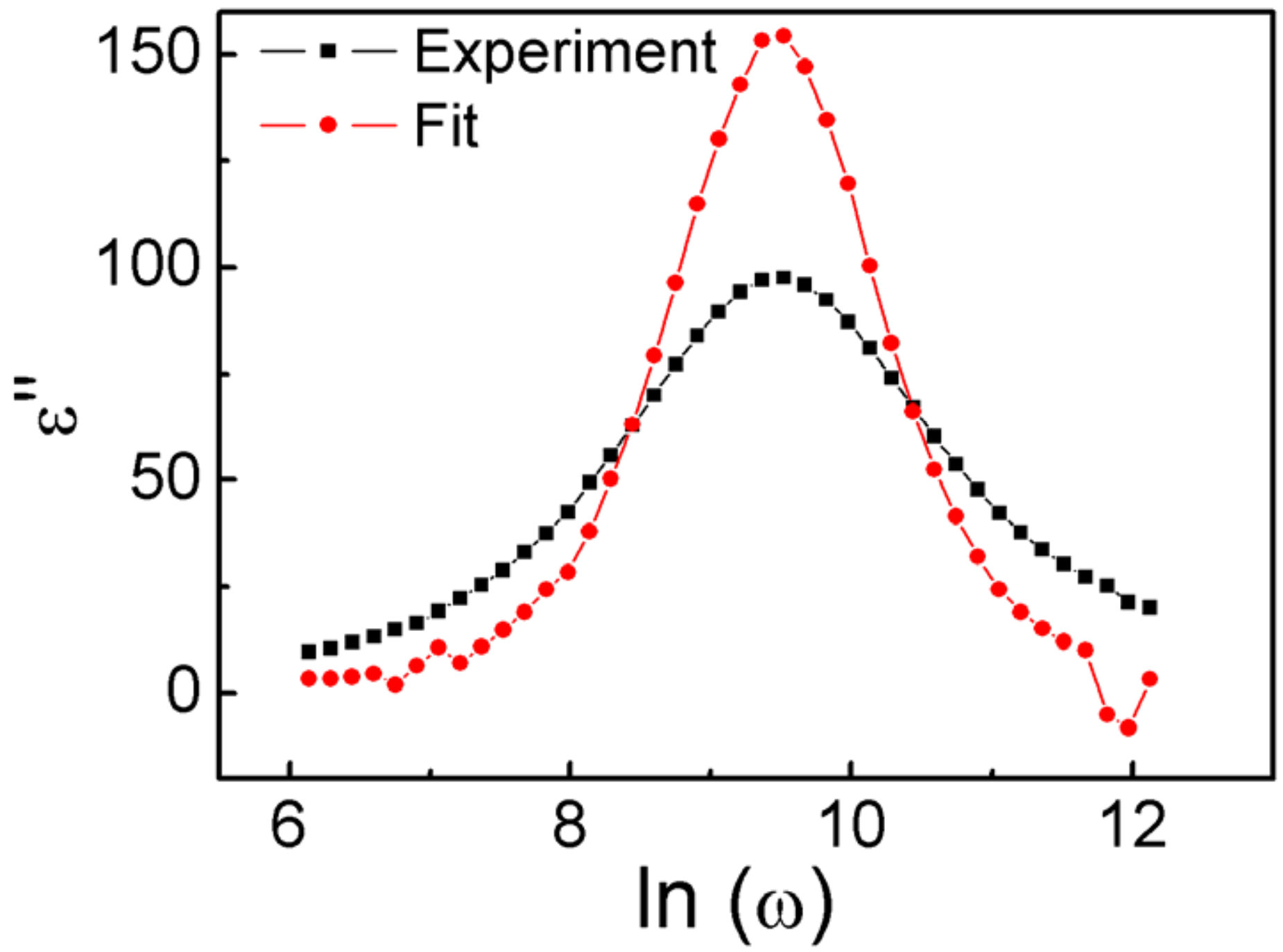

**FIG. 3.**

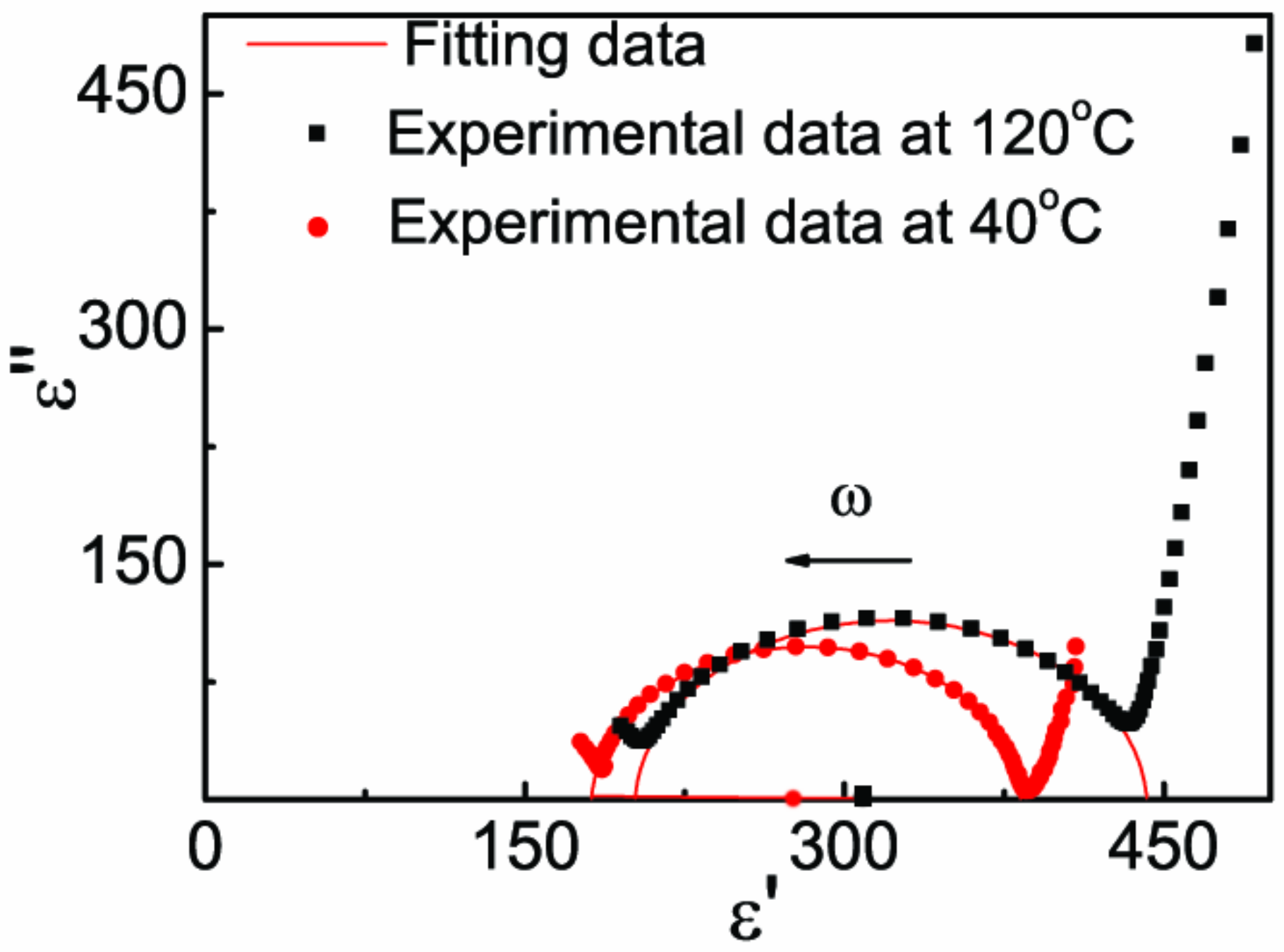

**FIG. 4.**

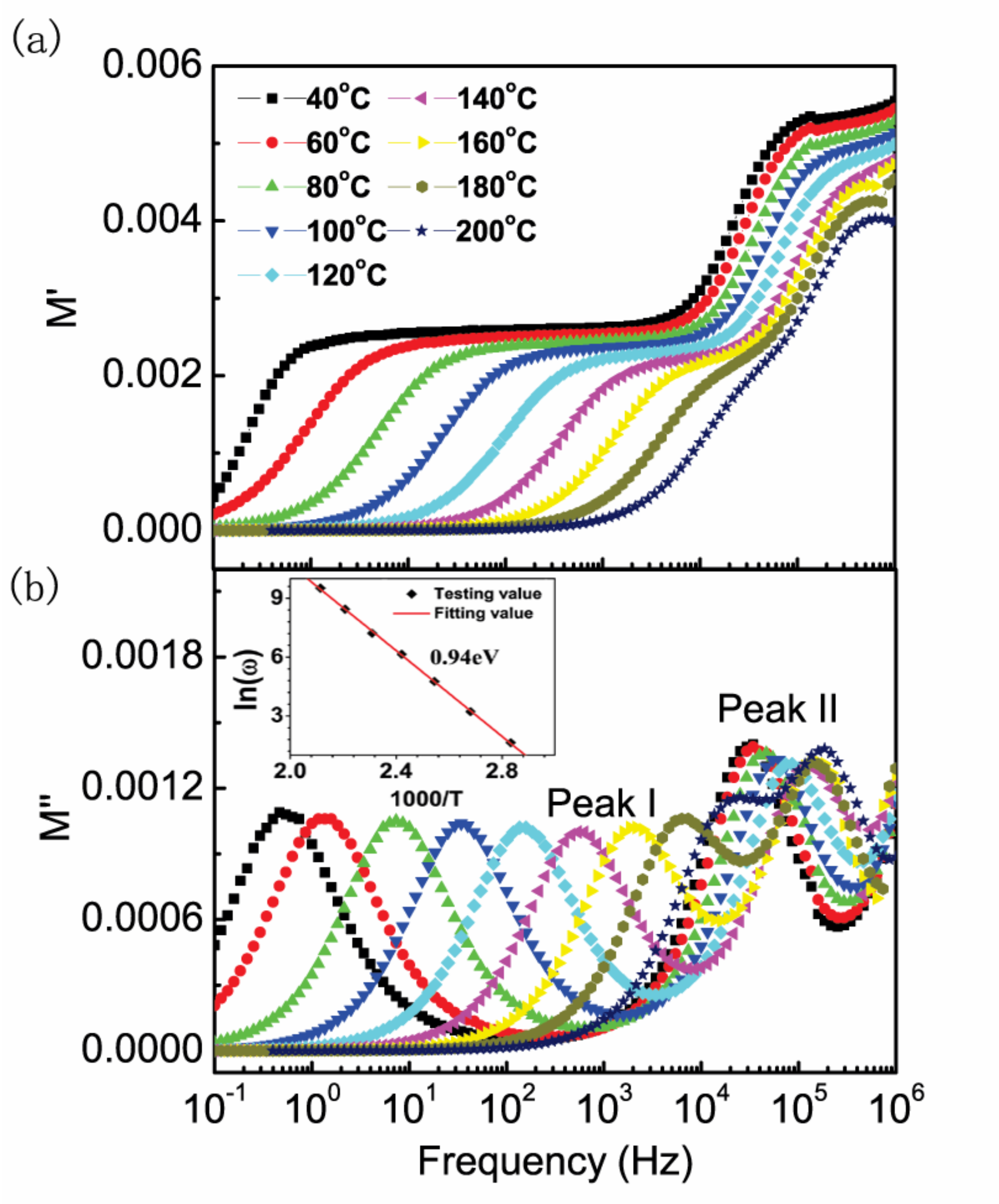

**FIG. 5.**

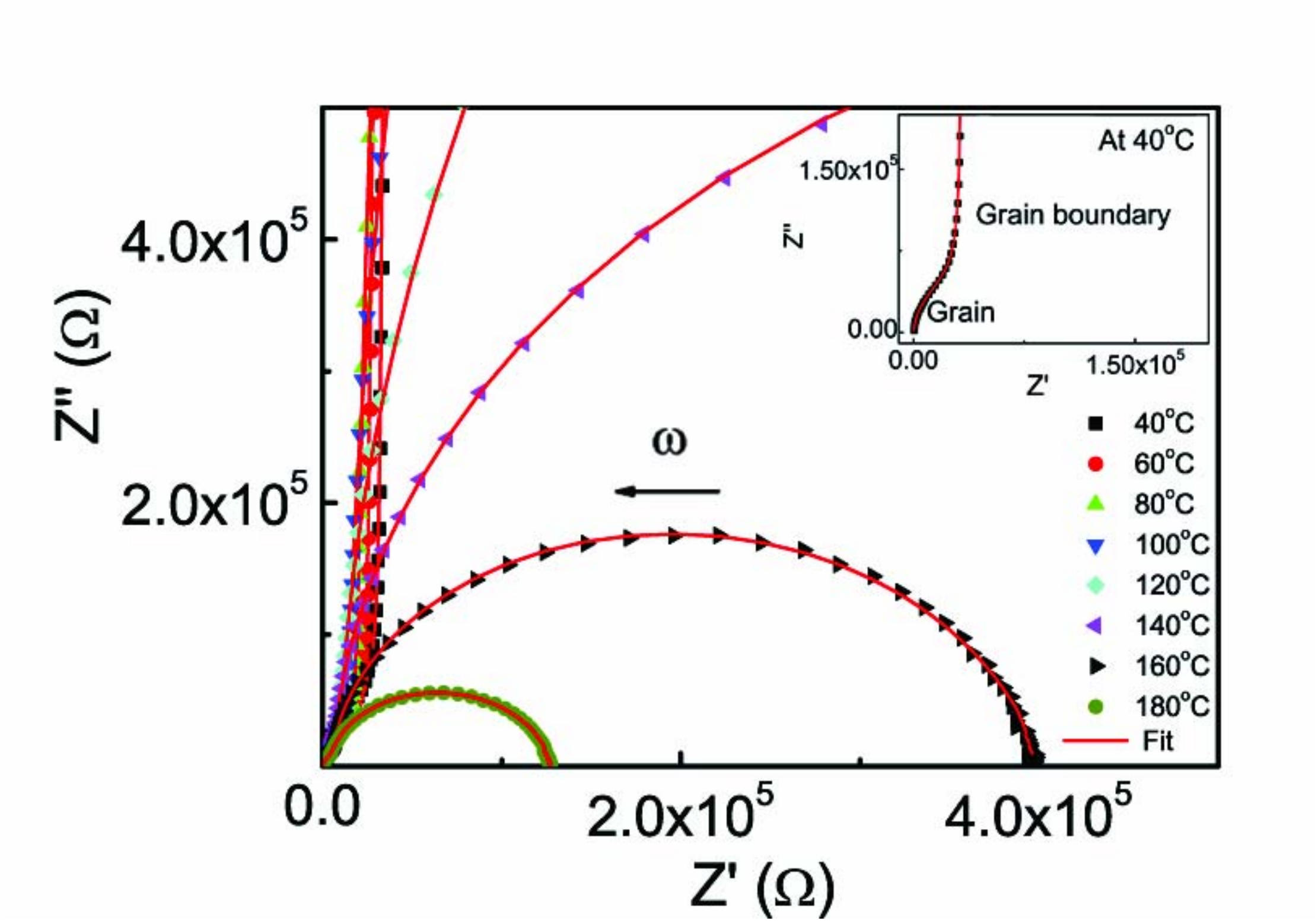

**FIG. 6**

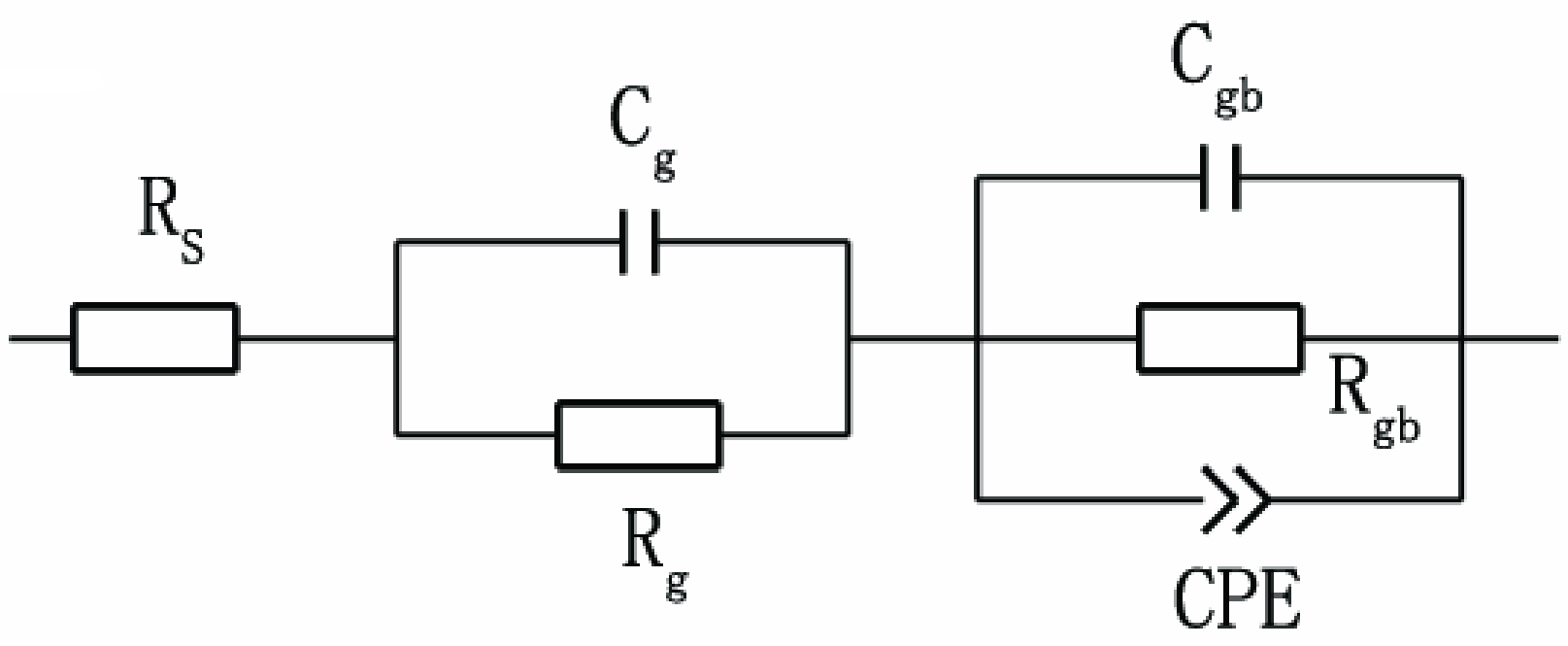

**FIG. 7.**

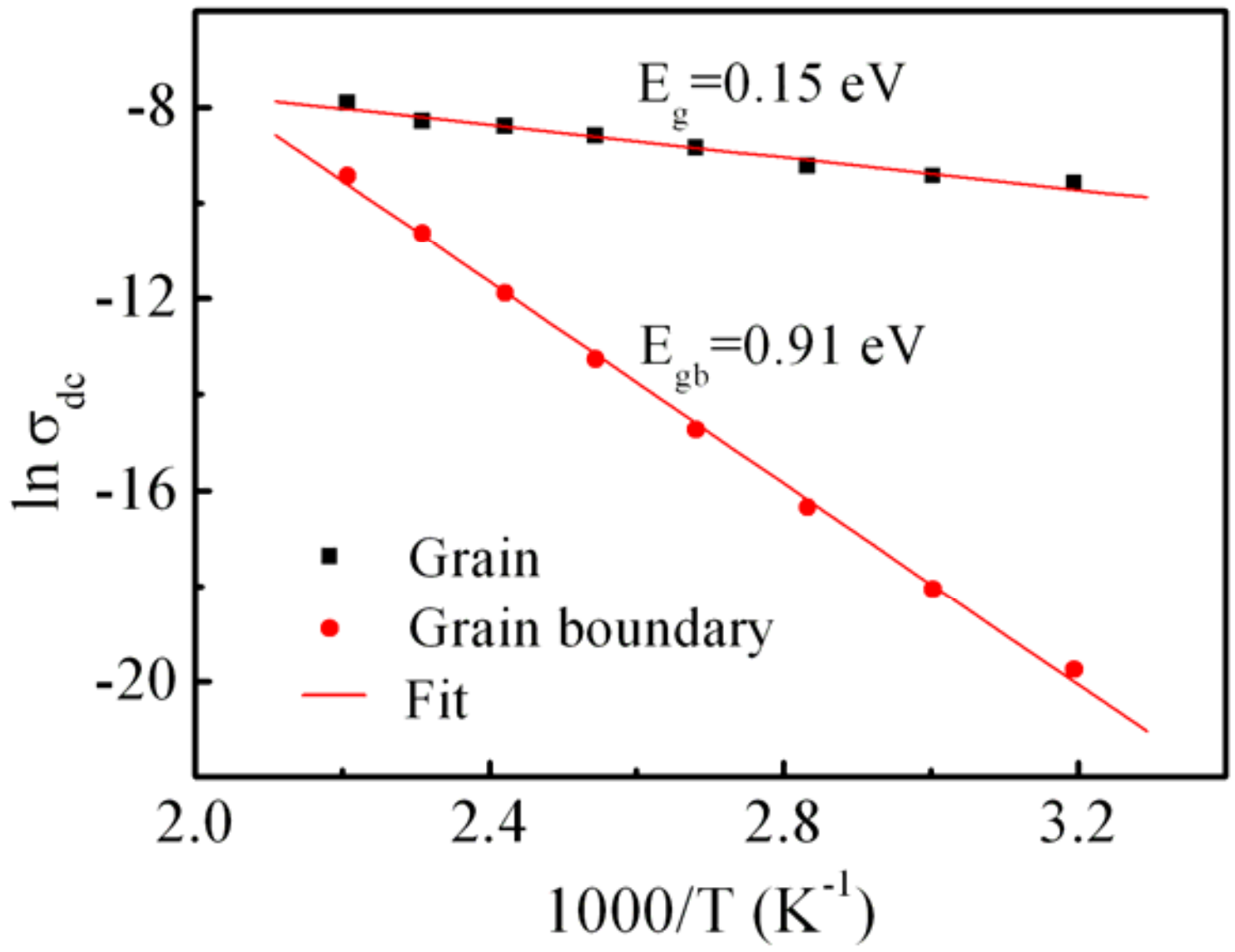

**FIG. 8.**

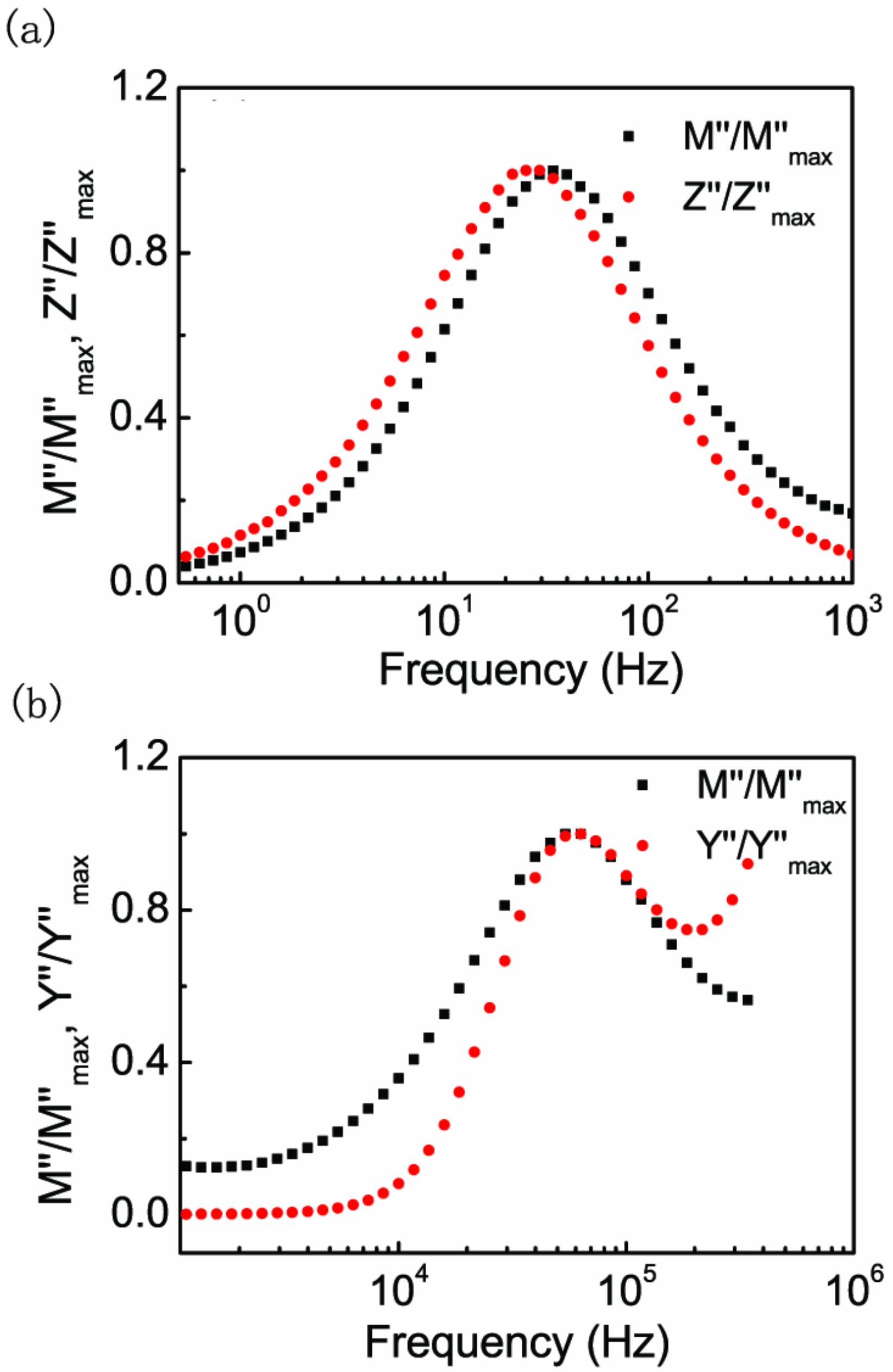

**FIG. 9.**

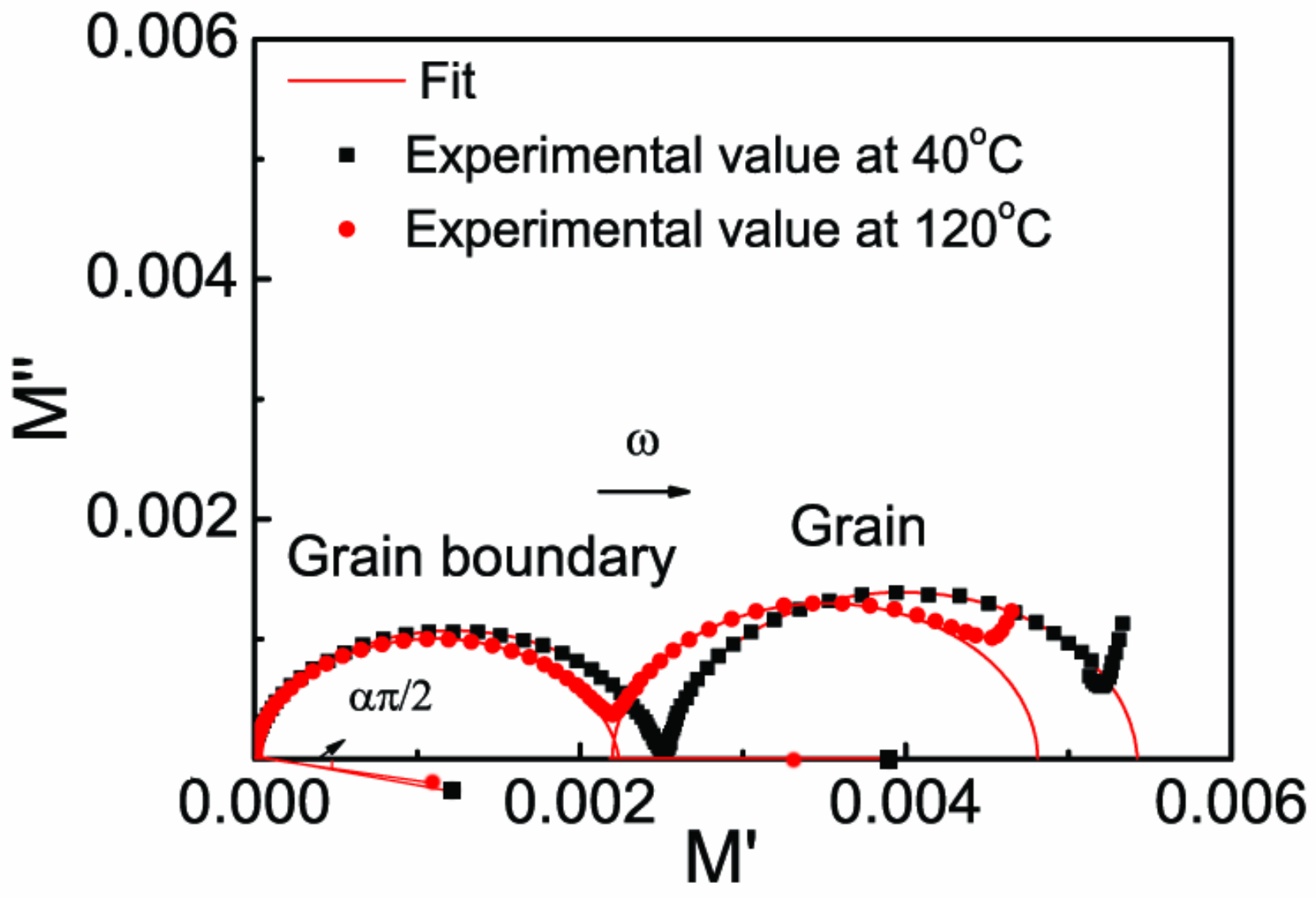

**FIG. 10**.

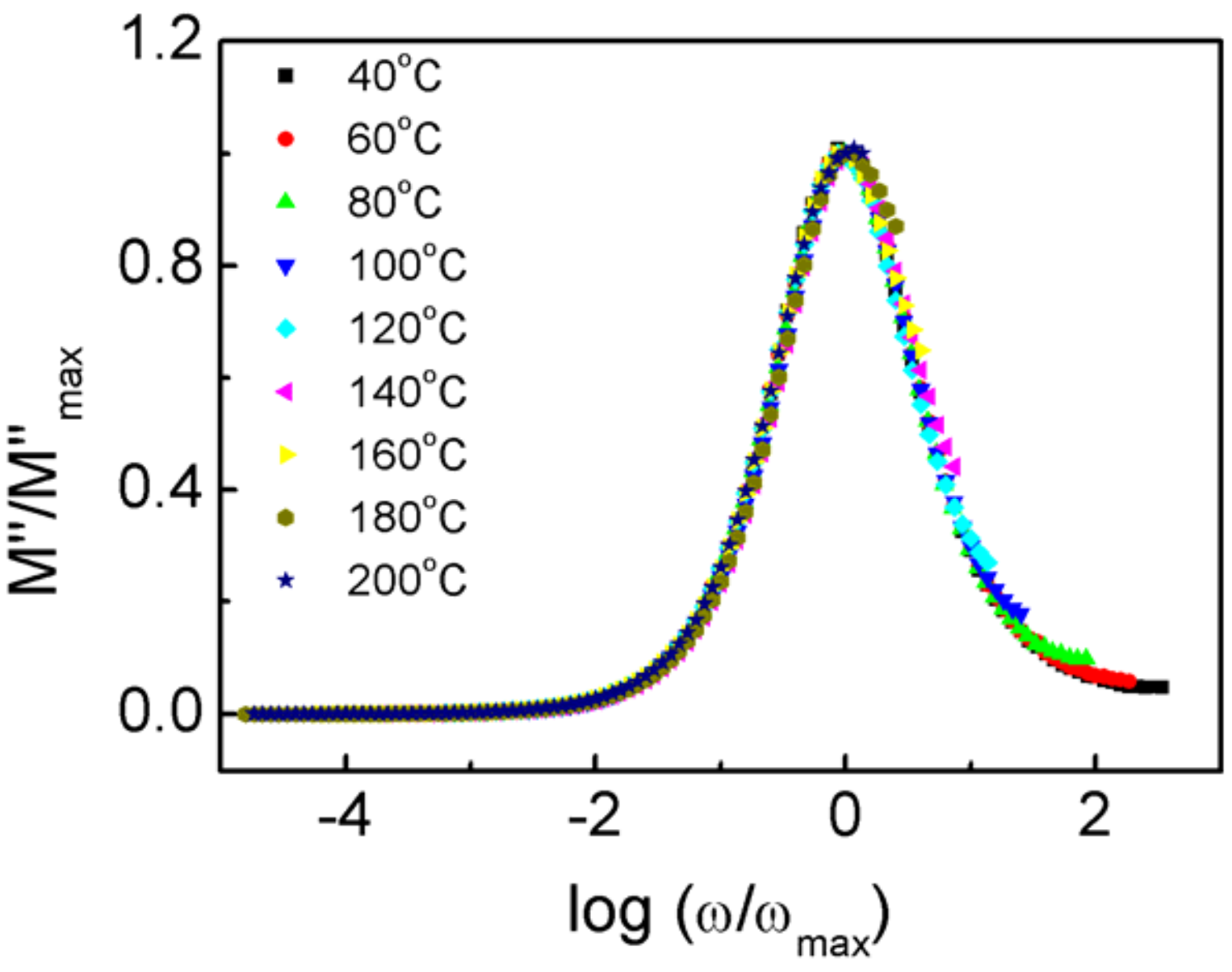